\documentclass[a4paper]{article}

\usepackage{INTERSPEECH2018}

\title{Attentive Statistics Pooling for Deep Speaker Embedding}
\name{Koji Okabe$^1$, Takafumi Koshinaka$^1$, Koichi Shinoda$^2$}
\address{
  $^1$Data Science Research Laboratories, NEC Corporation, Japan\\
  $^2$Department of Computer Science, Tokyo Institute of Technology, Japan}
\email{k-okabe@bx.jp.nec.com, koshinak@ap.jp.nec.com, shinoda@c.titech.ac.jp}

\begin{document}

\maketitle
\begin{abstract}
%
%

This paper proposes attentive statistics pooling for deep speaker embedding 
in text-independent speaker verification.
In conventional speaker embedding, frame-level features are averaged
over all the frames of a single utterance to form an utterance-level feature.
Our method utilizes an attention mechanism to give
different weights to different frames and generates not only
weighted means but also weighted standard deviations.
In this way, it can capture long-term variations in speaker characteristics more effectively. 
An evaluation on the NIST SRE 2012 and the VoxCeleb data sets shows that
it reduces equal error rates (EERs) from the conventional method by 7.5\% and 8.1\%, respectively.

 
\end{abstract}
\noindent\textbf{Index Terms}: speaker recognition, deep neural networks, attention, statistics pooling


\section{Introduction}
%
%

Speaker recognition
has advanced considerably in the last decade with the i-vector paradigm \cite{dehak2011front}, 
in which a speech utterance or a speaker is represented in the form of a fixed- 
low-dimensional feature vector.


With the great success of deep learning over a wide range of machine learning tasks, including automatic speech recognition (ASR), 
an increasing number of research studies have introduced deep learning into feature extraction for speaker recognition.  
In early studies \cite{lei2014novel, mclaren2015advances}, deep neural networks (DNNs) derived from acoustic models for ASR 
have been employed as a universal background model (UBM) to provide phoneme posteriors as well as bottleneck features, 
which are used for, respectively, zeroth- and first-order statistics in i-vector extraction.  
While they have shown better performance than conventional UBMs based on Gaussian mixture models (GMMs), 
they have the drawback of language dependency \cite{zheng2015exploring} and also require expensive phonetic transcriptions for training \cite{tian2016improving}.


Recently, DNNs have been shown to be useful for extracting speaker-discriminative feature vectors independently from the i-vector framework.  
With the help of large-scale training data, such approaches lead to better results, particularly under conditions of short-duration utterances. 
In fixed-phrase text-dependent speaker verification, an end-to-end neural network-based method has been proposed \cite{heigold2016end}
in which Long Short-Term Memory (LSTM) with a single output from the last frame is used to obtain utterance-level speaker features, 
and it has outperformed conventional i-vector extraction.


In text-independent speaker verification, where input utterances can have variable phrases and lengths, 
an average pooling layer has been introduced to aggregate frame-level speaker feature vectors 
to obtain an utterance-level feature vector, i.e., speaker embedding, with a fixed number of dimensions. 
Most recent studies have shown that DNNs achieve better accuracy than do i-vectors \cite{nagrani2017voxceleb, li2017deep}.  
Snyder \textit{et al.} \cite{snyder2017deep} employed an extension of average pooling, in which what they called statistics pooling 
calculated not only the mean, but also the standard deviation of frame-level features.
They, however, have not yet reported the effictiveness of standard deviaion pooling to accuracy improvement.



Other recent studies conducted from a different perspective \cite{bhattacharya2017deep, chowdhury2017attention}
have incorporated attention mechanisms \cite{raffel2015feed}.
It had previously produced significant improvement in machine translation.
In the scenario of speaker recognition, an importance metric is
computed by the small attention network that works as a part of the speaker embedding network.
The importance is utilized for calculating the weighted mean of frame-level feature vectors.
This mechanism enables speaker embedding to be focused on important frames
and to obtain long-term speaker representation with higher discriminative power. 
Such previous work, however, has been evaluated only in such limited tasks as fixed-duration text-independent \cite{bhattacharya2017deep} 
or text-dependent speaker recognition \cite{chowdhury2017attention}.


In this paper, we propose a new pooling method, called attentive statistics pooling, that provides 
importance-weighted standard deviations as well as the weighted means of frame-level features, 
for which the importance is calculated by an attention mechanism.
This enables speaker embedding to more accurately and efficiently capture speaker factors with respect to long-term variations.
To the best of our knowledge, this is the first attempt reported in the literature 
to use attentive statistics pooling in text-independent and variable-duration scenarios.
We have also experimentally shown, through comparisons of various pooling layers, the effectiveness of long-term speaker characteristics
derived from standard deviations.


The remainder of this paper is organized as follows:
Section 2 describes a conventional method for extracting deep speaker embedding.
Section 3 reviews two extensions for the conventional method, and then introduces the proposed speaker embedding method.
The experimental setup and results are presented in Section 4.
Section 5 summarizes our work and notes future plans.

\section{Deep speaker embedding}

\begin{figure}[t]
  \centering
  \includegraphics[width=\linewidth]{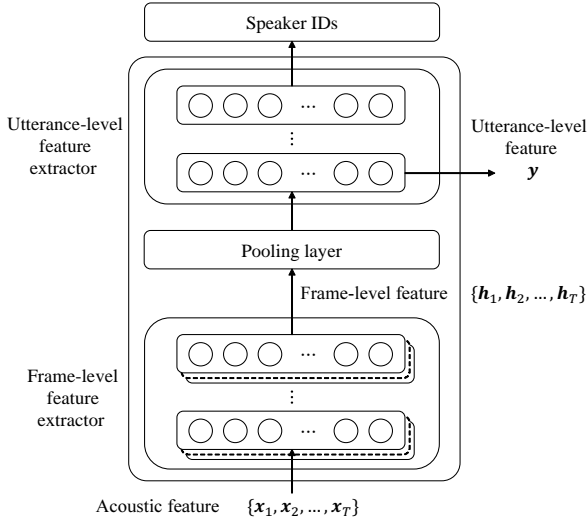}
  \caption{DNNs for extracting utterance-level speaker features}
  \label{fig:dnn}
\end{figure}

The conventional DNN for extracting utterance-level speaker features consists of three blocks, as shown in Figure~\ref{fig:dnn}.

The first block is a frame-level feature extractor. The input to this block is a sequence of acoustic features,
e.g., MFCCs and filter-bank coefficients. After considering relatively short-term acoustic features,
this block outputs frame-level features. 
Any type of neural network is applicable for the extractor, e.g.,
a Time-Delay Neural Network (TDNN) \cite{snyder2017deep}, 
Convolutional Neural Network (CNN) \cite{nagrani2017voxceleb, li2017deep},
LSTM \cite{bhattacharya2017deep, chowdhury2017attention}, 
or Gated Recurrent Unit (GRU) \cite{li2017deep}.

The second block is a pooling layer that converts variable-length frame-level features
into a fixed-dimensional vector.  The most standard type of pooling layer obtains the average of all frame-level features (average pooling).

The third block is an utterance-level feature extractor in which a number of fully-connected hidden layers
are stacked.
One of these hidden layers is often designed to have a smaller number of units (i.e., to be a bottleneck layer),
which forces the information brought from the preceding layer into a low-dimensional representation. 
The output is a softmax layer, and each of its output nodes corresponds to one speaker ID.
For training, we employ back-propagation with cross entropy loss.
We can then use bottleneck features in the third block as utterance-level features.
Some studies refrain from using softmax layers and achieve end-to-end neural networks 
by using contrastive loss \cite{nagrani2017voxceleb} or triplet loss \cite{li2017deep}.
Probabilistic linear discriminant analysis (PLDA) \cite{ioffe2006probabilistic, prince2007probabilistic} 
can also be used for measuring the distance between two utterance-level features \cite{snyder2017deep, bhattacharya2017deep}.

\section{Higher-order pooling with attention}

The conventional speaker embedding described in the previous section suggests the addition of two extensions of the pooling method:
the use of higher-order statistics and the use of attention mechanisms. 
In this section we review both and then introduce our proposed pooling method, which we refer to as attentive statistics pooling.

\subsection{Statistics pooling}

The statistics pooling layer \cite{snyder2017deep} calculates the mean vector $\bm{\mu}$
as well as the second-order statistics as the standard deviation vector $\bm{\sigma}$
over frame-level features $\bm{h}_{t}$ ($t=1, \cdots, T$). 
\begin{equation}
  \bm{\mu} = \frac{1}{T}\sum_{t}^{T}\bm{h}_{t} ,
  \label{eq1}
\end{equation}
\begin{equation}
  \bm{\sigma} = \sqrt{\frac{1}{T}\sum_{t}^{T}\bm{h}_{t}\odot\bm{h}_{t} - \bm{\mu}\odot\bm{\mu}} ,
  \label{eq2}
\end{equation}
where $\odot$ represents the Hadamard product.
The mean vector (\ref{eq1}) which aggregates frame-level features can be viewed as the main body of utterance-level features.
We consider that the standard deviation (\ref{eq2}) also plays an important role
since it contains other speaker characteristics in terms of temporal variability over long contexts.
LSTM is capable of taking relatively long contexts into account, using its recurrent connections and gating functions.
However, the scope of LSTM is actually 
no more than a second ($\sim$100 frames) due to the vanishing gradient problem \cite{bhattacharya2016deep}. 
A standard deviation, which is potentially capable of revealing any distance in a context, 
can help speaker embedding capture long-term variability over an utterance.

\subsection{Attention mechanism}

It is often the case that frame-level features of some frames are more unique and important 
for discriminating speakers than others in a given utterance.
Recent studies \cite{bhattacharya2017deep, chowdhury2017attention} have applied attention mechanisms 
to speaker recognition for the purpose of frame selection by automatically calculating the importance of each frame.

An attention model works in conjunction with the original DNN and calculates a scalar score $e_{t}$ for each frame-level feature
\begin{equation}
  e_{t} = \bm{v}^{T}f(\bm{W}\bm{h}_{t} + \bm{b}) + k ,
  \label{eq3}
\end{equation}
where
$f(\cdot)$ is a non-linear activation function, such as a tanh or ReLU function.
The score is normalized over all frames by a softmax function so as to add up to the following unity:
\begin{equation}
  \alpha_{t} = \frac{\exp(e_{t})}{\sum_{\tau}^{T}\exp(e_{\tau})} .
  \label{eq4}
\end{equation}

The normalized score $\alpha_t$ is then used as the weight
in the pooling layer to calculate the weighted mean vector
\begin{equation}
  \tilde{\bm{\mu}} = \sum_{t}^{T}\alpha_{t}\bm{h}_{t} .
  \label{eq5}
\end{equation}

In this way, an utterance-level feature extracted from a weighted mean vector focuses on 
important frames and hence becomes more speaker discriminative.

\subsection{Attentive statistics pooling}

The authors believe that both higher-order statistics (standard deviations as utterance-level features) and attention mechanisms are effective for
higher speaker discriminability. Hence, it would make sense to consider a new pooling method,
attentive statistics pooling, which produces both means and 
standard deviations with importance weighting by means of attention, as illustrated in Figure~\ref{fig:attentive_statistics_pooling}.
Here the weighted mean is given by (\ref{eq5}), and the weighted standard deviation is defined as follows:

\begin{figure}[t]
  \centering
  \includegraphics[width=\linewidth]{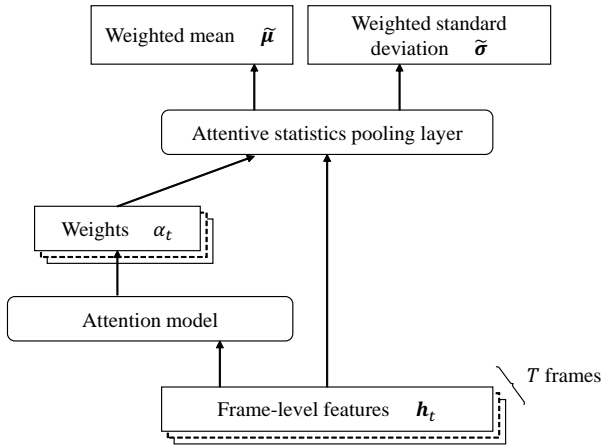}
  \caption{Attentive statistics pooling}
  \label{fig:attentive_statistics_pooling}
\end{figure}

\begin{equation}
  \tilde{\bm{\sigma}} = \sqrt{\sum_{t}^{T}\alpha_{t}\bm{h}_{t}\odot\bm{h}_{t} - \tilde{\bm{\mu}}\odot\tilde{\bm{\mu}}} ,
  \label{eq6}
\end{equation}
where the weight $\alpha_{t}$ calculated by (\ref{eq4}) is shared by both the weighted mean $\tilde{\bm{\mu}}$ and weighted standard deviation $\tilde{\bm{\sigma}}$.
The weighted standard deviation is thought to take the advantage of both statistics pooling and attention,
i.e., feature representation in terms of long-term variations and 
frame selection in accord with importance,
bringing higher speaker discriminability to utterance-level features.
Needless to say, as (\ref{eq6}) is differentiable, DNNs with attentive statistics pooling can be trained on the basis of back-propagation.

\section{Experiments}

\subsection{Experimental settings}

We report here speaker verification accuracy w.r.t. the NIST SRE 2012 \cite{greenberg20132012} Common Condition 2 (SRE12 CC2) 
and VoxCeleb corpora \cite{nagrani2017voxceleb}.
Deep speaker embedding with our attentive statistics pooling is compared to that with conventional statistics pooling and with attentive average pooling, as well as
with traditional i-vector extraction based on GMM-UBM.

\subsubsection{i-vector system}

The baseline i-vector system uses 20-dimensional MFCCs for every 10ms. 
Their delta and delta-delta features were appended to form 60-dimensional acoustic features.
Sliding mean normalization with a 3-second window and energy-based voice activity detection (VAD) were then applied, in that order.
An i-vector of 400 dimensions was then extracted from the acoustic feature vectors,
using a 2048-mixture UBM and a total variability matrix (TVM). 
Mean subtraction, whitening, and length normalization \cite{garcia2011analysis} were applied to the i-vector 
as pre-processing steps before sending it to the PLDA, and similarity was then
evaluated using a PLDA model with a speaker space of 400 dimensions.


\subsubsection{Deep speaker embedding system}


We used 20-dimensional MFCCs for SRE12 evaluation, 
and 40-dimensional MFCCs for VoxCeleb evaluation, for every 10ms.
Sliding mean normalization with a 3-second window and energy-based VAD were then applied in the same way
as was done with the i-vector system.

The network structure, except for its input dimensions, was exactly the same as the one shown in the recipe \footnote{egs/sre16/v2} 
published in Kaldi's official repository \cite{povey2011kaldi, snyder2018vector}.
A 5-layer TDNN with ReLU followed by batch normalization was used for extracting frame-level features.
The number of hidden nodes in each hidden layer was 512. The dimension of a frame-level feature for pooling was 1500.
Each frame-level feature was generated from a 15-frame context of acoustic feature vectors.

Pooling layer aggregates frame-level features, followed by 2 fully-connected layers with ReLU activation functions, batch normalization, and a softmax output layer.
The 512-dimensional bottleneck features from the first fully-connected layer were used as speaker embedding.

We tried four pooling techniques to evaluate the effectiveness of the proposed method:
(i) simple average pooling to produce means only, 
(ii) statistics pooling to produce means and standard deviations, 
(iii) attentive average pooling to produce weighted means, 
and (iv) our proposed attentive statistics pooling.
We used ReLU followed by batch normalization for activation function $f\left(\cdot\right)$ in (\ref{eq3}) of the attention model. 
The number of hidden nodes was 64.

Mean subtraction, whitening, and length normalization were applied to the speaker embedding, 
as pre-processing steps before sending it to the PLDA, and similarity was then
evaluated using a PLDA model with a speaker space of 512 dimensions.

\subsubsection{Training and evaluation data}

In order to avoid condition mismatch, different training data were used for each evaluation task w.r.t. SRE12 CC2 and VoxCeleb.


For SRE12 evaluation, telephone recordings from SRE04-10, Switchboard, and Fisher English were used as training data. 
We also applied data augmentation to the training set in the following ways: 
(a) Additive noise: each segment was mixed with one of the noise samples in the PRISM corpus \cite{ferrer2011promoting} (SNR: 8, 15, or 20dB), 
(b) Reverberation: each segment was convolved with one of the room impulse responses in the REVERB challenge data \cite{kinoshita2016summary}, 
(c) Speech encoding: each segment was encoded with AMR codec (6.7 or 4.75 kbps).
The evaluation set we used was SRE12 Common Condition 2 (CC2), 
which is known as a typical subset of telephone conversations without added noise.

For VoxCeleb evaluation, the development and test sets defined in \cite{nagrani2017voxceleb} were respectively used as training data and evaluation data.
The number of speakers in the training and evaluation sets were 1,206 and 40, respectively.
The number of segments in training and evaluation sets were 140,286 and 4,772, respectively.
Note that these numbers are slightly smaller than those reported in \cite{nagrani2017voxceleb} due to a few dead links on the official download server.
We also used the data augmentation (a) and (b) mentioned above.

We report here results in terms of equal error rate (EER) and the minimum of the normalized detection cost function,
for which we assume a prior target probability $P_{tar}$ of 0.01 ($\text{DCF10}^\text{-2}$) 
or 0.001  ($\text{DCF10}^\text{-3}$), and equal weights of 1.0 between misses $C_{miss}$ and
false alarms $C_{fa}$.

\subsection{Results}

\subsubsection{NIST SRE 2012}

Table~\ref{tab:sre12_result} shows the performance on NIST SRE12 CC2.
In the ``Embedding" column, {\it average} \cite{nagrani2017voxceleb, li2017deep} denotes average pooling that used only means, {\it attention} \cite{bhattacharya2017deep, chowdhury2017attention} used weighted means scaled by attention (attentive average pooling),
{\it statistics} \cite{snyder2017deep} used both means and standard deviations (statistics pooling), and {\it attentive statistics} is the proposed method (attentive statistics pooling), which used
both weighted means and weighted standard deviations scaled by attention.

\begin{table}[t]
  \caption{Performance on NIST SRE 2012 Common Condition 2. \textbf{Boldface} denotes the best performance for each column.}
  \label{tab:sre12_result}
  \centering
  \begin{tabular}{ l c c c}
    \toprule
    \textbf{Embedding} & $\textbf{DCF10}^\textbf{-2}$ & $\textbf{DCF10}^\textbf{-3}$ & \textbf{EER (\%)} \\

    \midrule
    i-vector              & \textbf{0.169}  & \textbf{0.291}  &         1.50   \\

    \midrule
    average   \cite{nagrani2017voxceleb, li2017deep}            &         0.290   &         0.484   &         2.57   \\
    attention  \cite{bhattacharya2017deep, chowdhury2017attention}           &         0.228   &         0.399   &         1.99   \\
    statistics \cite{snyder2017deep}           &         0.183   &         0.331   &         1.58   \\
    attentive statistics  &         0.170   &         0.309   & \textbf{1.47}  \\

    \bottomrule
  \end{tabular}
\end{table}

In comparison to average pooling, which used only means, the addition of attention was superior in terms of all evaluation measures.
Surprisingly, the addition of standard deviations was even more effective than that of attention. This indicates that long-term information
is quite important in text-independent speaker verification.
Further, the proposed attentive statistics pooling resulted in the best EER as well as minDCFs.
In terms of EER, it was 7.5\% better than statistics pooling.
This reflects the effect of using both long-context and frame-importance.
The traditional i-vector systems, however, performed better than speaker embedding-based systems, except for performance w.r.t. EER.
This seems to have been because the SRE12 CC2 task consisted of long-utterance trials
in which durations of test utterances were from 30 seconds to 300 seconds
and durations of multi-enrollment utterances were longer than 300 seconds.

Table~\ref{tab:sre12_duration} shows comparisons of EERs for several durations on NIST SRE12 CC2.
We can see that deep speaker embedding offered robustness in short-duration trials.
Although i-vector offered the best performance under the longest-duration condition (300s), 
our attentive statistics pooling achieved the best under all other conditions, 
with better error rates than those of statistics pooling under all conditions, including Pool (overall average).
Only attentive statistics pooling showed better performance than i-vectors on both 30-second trials and 100-second trials.

\begin{table}[t]
  \caption{EERs (\%) for each duration on NIST SRE 2012 Common Condition 2. \textbf{Boldface} denotes the best performance for each column.}
  \label{tab:sre12_duration}
  \centering
  \begin{tabular}{ l c c c c}
    \toprule
    \textbf{Embedding} & \textbf{30s} & \textbf{100s} & \textbf{300s} & \textbf{Pool} \\

    \midrule
    i-vector              &         2.66   &         1.09    & \textbf{0.58} &         1.50   \\

    \midrule
    average \cite{nagrani2017voxceleb, li2017deep}              &         3.58   &         2.07    &         1.86  &         2.57   \\
    attention \cite{bhattacharya2017deep, chowdhury2017attention}            &         3.00   &         1.58    &         1.27  &         1.99   \\
    statistics \cite{snyder2017deep}           &         2.49   &         1.25    &         0.82  &         1.58   \\
    attentive statistics  & \textbf{2.46}  & \textbf{1.07}   &         0.80  & \textbf{1.47}  \\

    \bottomrule
  \end{tabular}
\end{table}

\subsubsection{VoxCeleb}

Table~\ref{tab:voxceleb_result} shows performance on the VoxCeleb test set.
Here, also, the addition of both attention and of standard deviations helped improve
performance. As in the SRE12 CC2 case, standard deviation addition had a larger impact
than that of attention. The proposed attentive statistics pooling achieved
the best performance in all evaluation measures
and was 8.1\% better in terms of EER than statistics pooling.
This may have been because the durations were shorter than those with SRE12 CC2 (about 8 seconds on average in the evaluation),
and speaker embedding outperformed i-vectors, as well.
It should be noted that compared to the baseline performance shown in \cite{nagrani2017voxceleb}, whose best EER was 7.8\%, 
our experimental systems achieved much better performance, even though we used slightly smaller training and evaluation sets 
due to lack of certain videos.

\begin{table}[t]
  \caption{Performance on VoxCeleb. \textbf{Boldface} denotes the best performance for each column.}
  \label{tab:voxceleb_result}
  \centering
  \begin{tabular}{ l c c c}
    \toprule
    \textbf{Embedding} & $\textbf{DCF10}^\textbf{-2}$ & $\textbf{DCF10}^\textbf{-3}$ & \textbf{EER (\%)} \\

    \midrule
    i-vector              &         0.479   &         0.595   &         5.39   \\

    \midrule
    average \cite{nagrani2017voxceleb, li2017deep}              &         0.464   &         0.550   &         4.70   \\
    attention \cite{bhattacharya2017deep, chowdhury2017attention}            &         0.443   &         0.598   &         4.52   \\
    statistics \cite{snyder2017deep}           &         0.413   &         0.530   &         4.19   \\
    attentive statistics  & \textbf{0.406}  & \textbf{0.513}  & \textbf{3.85}  \\

    \bottomrule
  \end{tabular}
\end{table}

\section{Summary and Future Work}

We have proposed attentive statistics pooling for extracting deep speaker embedding.
The proposed pooling layer calculates weighted means and weighted standard
deviations over frame-level features scaled by an attention model.
This enables speaker embedding to focus only on important frames.
Moreover, long-term variations can be obtained as speaker characteristics in the standard deviations.
Such a combination of attention and standard deviations produces a synergetic effect to give deep speaker embedding higher discriminative power.
Text-independent speaker verification experiments on NIST SRE 2012 and VoxCeleb evaluation sets
showed that it reduced EERs from a conventional method by, respectively, 7.5\% and 8.1\% for the two sets.
While we have achieved considerable improvement under both short- and long-duration conditions, 
i-vectors are still competitive for long durations (e.g., 300s in SRE12 CC2). 
Pursuing even better accuracy under such conditions is an issue for our future work.



\bibliographystyle{IEEEtran}

\bibliography{okabebib}

\end{document}